\documentclass[aps,showpacs,twocolumn,preprintnumbers,notitlepage, secnumarabic,amsmath,amssymb,10pt, nofootinbib]{revtex4-2}
\usepackage{dcolumn}
\usepackage{bm}         
\usepackage{graphicx}
\usepackage{xcolor}
\usepackage{amsmath,amssymb}  
\usepackage{bm}  
\usepackage[pdftex,bookmarks,colorlinks,breaklinks]{hyperref}  
\begin{document}
\title{Dynamics of the transitions epochs in cosmological evolution}
\author{Bob Osano$^{1,2}$ \\
\small{$^{1}$Centre for Higher Education Development,\\ }$\&$\small{\\$^{2}$Cosmology and Gravity Group, Department of Mathematics and Applied Mathematics,\\ University of Cape Town (UCT), Rondebosch 7701, Cape Town, South Africa\\}
}
\email{bob.osano@uct.ac.za}

\begin{abstract} 

We study two transition periods in cosmology: radiation-to-matter and matter-to-dark energy. In each case, we define a new parameter $\chi$ given by the ratios of the two energy densities involved in the transition. Our study of the second epoch is motivated by the need to understand cosmic acceleration. Assuming a dynamic dark energy is the driving force for cosmic acceleration, we formulate a new equation of state for the dark energy given in terms of the ratio $\chi$ and the deceleration parameter, $q$. We have analysed the resultant system of equations, where we vary different parameters and examine the effect on the universe's evolution. For cosmic acceleration to occur, the EoS of the dynamic dark energy must lie in the interval  $\omega_{DDE}<-2/3$ at matter-dynamic dark energy equality( equivalently $\omega_{DDE}<-0.47$ today)   \end{abstract}
\date{\today}
\maketitle
\section{introduction}
 Cosmic acceleration, which is now fully confirmed \cite{Wein13,Alb06,Riess98}, is both confounding and vexing. Cosmological studies were heading towards settling on the $ \Lambda $CDM as the best-fit model for the universe. What appeared to remain was the precise determination of cosmological parameters; $h$, $\Omega_{m}$, $\Omega_{b}$, $\Omega_{\Lambda}$, $\Omega_{r}$, $\Omega_{\nu}$, $\Delta^{2}_{\mathcal{R}}(k_{*})$, $n$, $r$ and $\tau$  \cite{lahav2004} which are the Hubble parameter, total matter density, baryon density, the cosmological constant, radiation density, neutrino density perturbation amplitude, density perturbation spectral index, tensor ration and the ionisation optical depth respectively. Efforts to establish these parameters using different measurements and observational data have sometimes given conflicting results or results that challenge the underlying model \cite{Hin13,Agh18,Riess18}. Cosmic acceleration and the Hubble tension are two examples. It is unclear what exact value the Hubble constant ( the current value of the Hubble parameter) takes given that different measurement methods generate different values \cite{Reiss24}. Despite these,  the analysis of the universe dominated by radiation earlier on in its history and then matter has given us a good understanding of certain periods in the evolution history of the universe. We, for example, now know that the universe transited from radiation-dominated when the coupling of baryons and photons did not allow the former to cluster owing to radiation pressure. This meant that perturbations in cold dark matter grew at a slow logarithmic rate. The result was that structure growth was impeded during the epoch of radiation domination \cite{Cunn22}. The transition to matter domination saw a cessation of the suppression of the growth of small-scale perturbation. This occurs at the matter-radiation equality. 
 The smaller mode of perturbation that entered the horizon early in the radiation domination experienced greater suppression in its growth. For this reason, the power spectrum is a decreasing function of $\kappa$ on small scales. The turnover forms at a scale corresponding to the horizon size at matter–radiation equality. We know that a greater abundance of matter changes the point at which matter–radiation equality occurs, meaning that the turnover feature is sensitive to matter density and other parameters. It is this reason which makes it a viable tool for probing cosmology. \cite{Eis98}\cite{Dod03}.
 The premise of this paper is the possibility of a recent turnover involving a transition between matter domination and dark energy domination. Since the accelerated expansion was discovered, investigations into dark energy as a potential driver for cosmic expansion have seen a surge as seen in \cite{Vag,Bir19,Kno19,Fin19,Mortonson13,Tsu11,Li11} and references therein, to mention but a few. 
 
 In light of what we know about the matter-radiation transition and its potential in cosmological probes, there is much to be gained by probing the transition to dark energy domination. We see this as invaluable in resolving some conundrums in cosmology.  We begin with the primary equations that form the basis of the rest of the paper.
 \section{\label{one}Friedmann Equations}
The hot big-bang cosmological model (\cite{Weinberg72,Peebles93}) is the preferred model of the universe. This is a mathematical description based on the isotropic and homogeneous $Friedmann-Lema\hat{i}tre-Robertson-Walker$ (FLRW) solution of Einstein's equation, where the expansion of the Universe is manifested in the cosmic scale factor $a(t)$\cite{Tian17}. The expansion of the universe is itself governed by the Friedmann equations which take the form:
\begin{eqnarray}\label{Frid1}
\frac{\dot{a}^2}{a^2}&=&\frac{8\pi G}{3}\rho_{Tot}-\frac{k}{a^2}+\frac{\Lambda}{3},\\\label{Frid2}
\frac{\ddot{a}}{a}&=&-\frac{4\pi G}{3}\left(\rho_{Tot}+3 p_{Tot}\right)+\frac{\Lambda }{3},
\end{eqnarray} where $G$ is the gravitation constant, $k$ is the curvature, $\rho_{Tot}$ the energy density, $p_{Tot}$ the isotropic pressure. We have set the speed of light, $c$, to 1.
Now since the Hubble parameter $H=\dot{a}/a$, equation (\ref{Frid1}) can be appropriately normalised to read
\begin{eqnarray}\label{eqn1}
1 &=&\frac{8\pi G}{3H^2}\rho_{Tot}-\frac{k}{a^2H^2}+\frac{\Lambda}{3H^2}.
\end{eqnarray} This provides a simple yet effective way to discuss the energy-density composition of the universe.
The density parameter, $\Omega$, \cite{Perlmutter98} is the ratio of the actual (or observed) density $\rho$ to the critical density $\rho_{c}$ of the FRLW universe. We know that the relation between the actual density and the critical density determines the universe's overall geometry, i.e. when the two are equal, the universe is Euclidean (flat). Critical density in earlier models, which did not include a cosmological constant term, was used to delineate open and closed models.

\section{\label{three}Interactions and the evolution of total energy density}
In this section we examine the evolution of the total energy density of an FLRW universe made up of the following: radiation ($\rho_{r}$), baryonic and ordinary matter ($\rho_{m}$), dark matter ($\rho_{DM}$). We also use the notation $\rho_{M} (=\rho_{DM}+\rho_{m})$; the aggregation of dark matter and baryonic matter. dynamic dark energy is denoted by $\rho_{DDE}$ and non-dynamic dark energy by $\rho_{NDE}$, this distinction is important as what is collectively referred to as dark energy may be composed of distinctive subparts. The dynamic dark energy is taken to interact with dark matter. The formulation is such that it is easy to switch off interaction if desired. The question to ask is what kinds of interactions take place and are these significant enough to affect how these densities evolve? Between radiation and matter, the interaction term may involve ionisation. Radioactive particles or electromagnetic waves that are sufficiently energetic collide with atoms thereby knocking off electrons. For partially ionised matter, the growth pattern of radiation, ionised matter, and neutral matter will differ from that of just radiation and matter. But is this sufficient to affect the evolution of $\rho_{Tot}$? 

In this study, we let the dark-sector constituents mimic a perfect fluid obeying a barotropic equation on state. The interactions considered are not viscous or dissipative and therefore do not need the extended thermodynamics theory \cite{Bob,Maartens96,Hiscook83,IsraelStewart79,Israel89}. A case of dissipative flow will be examined elsewhere \cite{Bob2024}. The energy-momentum tensor in the present case takes the form
\begin{eqnarray}\label{Ten0}T_{Tot}^{\nu\mu}=(\rho_{Tot}+p_{Tot})u_{\nu}u_{\mu}+p_{Tot},g_{\nu\mu}\end{eqnarray} which obey the conservation law, $\nabla^{\nu}T_{Tot}^{\nu\mu}=0$. individual energy densities making up $\rho_{Tot}$ obey their evolution equations but may couple to others via interaction. We will include hypothetical interaction terms between pairs of equations. These are judiciously chosen in line with epochs and the transition periods between epochs. In particular, the evolutionary history indicates periods of transition from radiation domination to matter domination. We might consider the interaction between radiation and matter and use $Q_{rm}$ to denote it. Likewise, we use $Q_{ME}$ to denote the interaction between dark matter and dark energy. Our analysis will ignore any potential interaction between dark matter and baryons \cite{Barkana2018}. The individual density evolution equations take the form:
 \begin{eqnarray}\label{rhodots1}
\dot{\rho}_{r}&=&-3H(1+\omega_{r})\rho_{r}+Q_{rm}\\\label{rhodots2}
\dot{\rho}_{m}&=&-3H(1+\omega_{m})\rho_{m}-Q_{rm}\\\label{rhodots3}
\dot{\rho}_{DM}&=&-3H(1+\omega_{DM})\rho_{DM}+Q_{ME}\\\label{rhodots4}
\dot{\rho}_{DDE}&=&-3H(1+\omega_{DDE})\rho_{DDE}-Q_{ME}.\\\label{rhodots5}
\dot{\rho}_{NDE}&=&0.
\end{eqnarray}
We are only interested in the transition dynamics and will proceed to review the relevant epochs. First, we must clarify what we mean by a matter type dominating the dynamics of a given epoch. 
\section{Dominant matter type}
Although domination requires one density to be higher than the rest, we choose the upper limit of this and demand that it is higher than all the remaining combined \cite{Bob21}. For the context of this analysis, a matter type is dominant if its energy density is $50\%$ of $\rho_{Tot}$ or higher.  In principle, this definition ensures that the dominant matter type drives the universe's expansion rate in the said epoch.\section{Radiation-Matter transition}
The transition from radiation domination to matter domination is characterised by the {\it radiation} and {\it matter} constituting the greater proportion of the cumulative energy density, and as previously mentioned, we interpret this as the dominant density which accounts for at least half of the total energy density. We will define a new parameter to study the transition period. To this end, we note that  subtracting equation (\ref{Frid1})from equation (\ref{Frid2}), setting $8\pi G=1$ and $\kappa=0$, and expressing the resultant equation in terms of the Hubble parameter using the system of equations (\ref{rhodots1}) yields
\begin{eqnarray}\label{def0}
\dot{H}&=&-\frac{1}{2}[(1+w_{r})\rho_{r}+(1+w_{m})\rho_{m})\nonumber\\
&&+(1+w_{DM})\rho_{DM})+(1+w_{DDE})\rho_{DDE}].
\end{eqnarray} $H$ has the dimension of time. A transition epoch involves the preceding and the succeeding dominant energy densities. For example, the radiation-to-matter transition has radiation as preceding and matter as succeeding. We can rewrite equation (\ref{def0}) by factoring out the preceding energy density such that
\begin{eqnarray}\label{def1}
\dot{H}&=&-\frac{1}{2}\rho_{r}\biggl[(1+w_{r})+(1+w_{m})\frac{\rho_{m}}{\rho_{r}})\nonumber\\
&&+(1+w_{DM})\frac{\rho_{DM}}{\rho_{r}})+(1+w_{DDE})\frac{\rho_{DDE}}{\rho_{r}}\biggr].
\end{eqnarray} The ratios of energy densities of the subdominant species are taken as small in comparison to $\rho_{M}/\rho_{r}$ and hence can be set to zero. In principle, the subdominant densities do not vanish but have a negligible contribution to the Hubble parameter (\ref{def1}) hence, 
\begin{eqnarray}\label{def2}
\dot{H}&\thickapprox&-\frac{1}{2}\rho_{r}[(1+w_{r})+(1+w_{m})\frac{\rho_{m}}{\rho_{r}})].
\end{eqnarray} It is the last term in equation (\ref{def2}) that gives a hint of how to define a new parameter.
\subsection{A density ratio parameter}
We define a new parameter given by the ratio:
\begin{eqnarray}\label{param1}
\chi_{rm}&=&\frac{\rho_{m}}{\rho_{r}}.
\end{eqnarray}  Although $0<\chi_{rm}<\infty$, we are only interested in the dynamics around matter-radiation equality, which in terms of the new parameter is around $\chi_{rm}=1$. We, for example, know that the horizon size sets the position of the peak of the matter power spectrum at the epoch of matter-radiation equality \cite{Bahr23}. The position of the first peak is consistent with a flat universe.

A variant of equation (\ref{def2}) is given in the dark matter - dark energy transition epoch section. For now, taking the time derivative of $\chi_{rm}$ and using equations (\ref{rhodots1} and \ref{rhodots2}) yields 
\begin{eqnarray}\label{chi3}
\dot{\chi}_{rm}&=&-\Theta(w_{m}-w_{r})\chi_{rm}-\frac{Q_{rm}}{\rho_{r}}(1+\chi_{rm}),
\end{eqnarray} where $w_{m}=0$ and $w_{r}=1/3$.

We shortly show that equation (\ref{chi3}) is coupled to other equations but before we get to that let us first discuss the interaction term in this equation (\ref{chi3}). The material whose densities are considered here may experience different interactions but not including those that may change the nature of interacting material (e.g. chemical). The effect of the interaction may be quantified by detailing the proportion of materials undergoing interaction. For this reason, the interaction term $Q_{rm}$ may be expressed as a proportion of the ratio $\chi_{rm}$. i.e. $Q_{rm}=\psi(t)\chi_{rm}$, where the highest interaction occurs at equality ( $\chi_{rm}=1$) for that is when the largest percentage of the two fluids may experience interaction. Even though the system we consider is nonlinear and not time-invariant, the product $\psi(t)\chi_{rm}$ mimics a transfer function. 
Equation (\ref{chi3}) takes the simple form:
\begin{eqnarray}\label{chi4}
\dot{\chi}_{rm}&=&-\Theta(w_{m}-w_{r})\chi_{rm}-\frac{\psi \chi_{rm}}{\rho_{r}}(1+\chi_{rm}).
\end{eqnarray} This is one of the main equations in the radiation-matter transition epoch. 
The other relevant equations are the evolution equation for radiation given by equation (\ref{rhodots2}) and equation (\ref{def2}). If we let $X=\chi_{rm}$ and $Z=1/\rho_{r}$ and apply $w_{r}=1/3$, $w_{m}=0$, we can express the system in the following compact form.
\begin{eqnarray}\label{sys01}
\dot{X}&=&H X-\psi X Z(1+X)\\\label{sys02}
\dot{Z}&=& 4H Z-\psi X Z^2\\\label{sys04}
\dot{H}&=&-\frac{1}{Z}(\frac{X}{2}+\frac{2}{3}).
\end{eqnarray} This gives a generic system of equations that allows us to study the transition from radiation to matter domination. In fact, in standard particle cosmology \cite{Sug95}, one comes across a similar ratio correlation of the form $1+\rho_{\nu}/\rho_{\gamma}$where $\rho_{\nu}$ is the density of neutrinos and $\rho_{\gamma}$ is photons density. 

\begin{figure}[h!]
\centering
\includegraphics[scale=0.45]{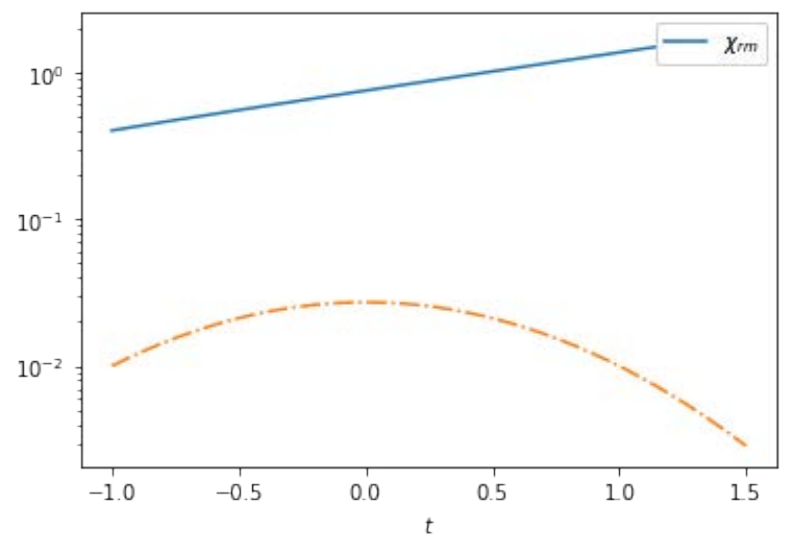}
\caption{\it This plot shows how $\chi_{rm}$ and $\psi$ grow with cosmic time. We note that the scale is normalised so that $t_{eq}=0$}
\label{fig3a}
\end{figure}

\begin{figure}[h!]
\centering
\includegraphics[scale=0.45]{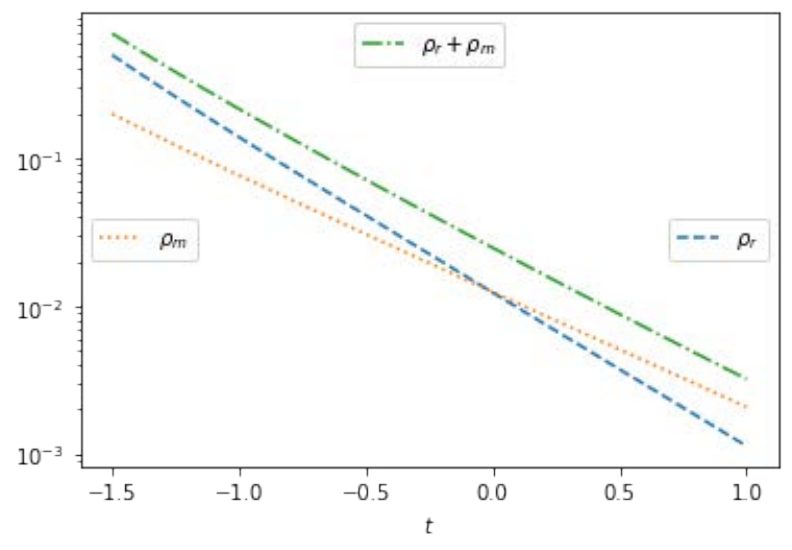}
\caption{\it The plot shows the devlopment of $\rho_{r}$, $\rho_{m}$ and the sum $\rho_{m}+\rho_{r}$ plotted against cosmic time. We have used a log scale on both axes. The plots are normalised such that $t_{eq}=0$ (equivalently $\chi_{rm}=1$). }\label{fig3b}
\end{figure}

We emphasise that the assumptions and ansatz used in obtaining the numerical solution to this system are arbitrary and chosen for illustration purposes. We need $\phi$ to be a cosmic time-dependent function with a maximum value at $t_{eq}$ or when $\chi_{rm}=1$. We use the ansatz $\psi(t)=e^{-t^2}$ where $t_{eq}=0$ is at matter-radiation equality. The following initial conditions are used in obtaining numerical solutions presented in Figures (\ref{fig3a}) and (\ref{fig3b}): $X_{rm}=0.4$, $Z=2$ and $H=0.67$.

\subsection{Results and Analysis}
We see that hypothetical interactions quantified here by $Q_{rm}$ affect the dynamics of the universe particularly near equality. In this study, we have deliberately limited the initial condition of the interaction term where we use an exponential ansatz as a driving force of the term. In this illustration, the EoS of the two competing densities are known. The primary parameter, $\chi_{rm}$, increases with cosmic time as seen in figure (\ref{fig3a}). The two densities grow as expected around $\chi_{rm}=1.$ The interaction terms play the role of changing the time to equality. In terms of physics, this would impact structure formation. We know that structures that are smaller than the horizon experience stunted growth during radiation domination. A delay in time to equality affects an even greater percentage of such structures. As the universe expands, the radiation density drops faster than matter as indicated in Figure (\ref{fig3b}). This is due to the redshifting of photon energy leading to an estimated crossover or matter-radiation equality ($\chi_{rm}=1$) at about 50,000 years after the Big Bang.  Beyond $\chi_{rm}=1$, fluctuations in all dark matter potentially grow unimpeded and create kernels into which the baryons can later fall. The turnover in the matter power spectrum is induced by the particle horizon at this epoch. This turnover can be measured in large redshift surveys \cite{Bahr23}.
The procedure and analysis in this section set the foundation for studying the matter-dark energy transition where the dynamic dark energy EoS is unknown.

\section{Matter-Dark energy transition}
From equation (\ref{def0}), we factor out the dark matter energy density and ignore density ratios that are negligible in comparison to $\rho_{DDE}/\rho_{M}$ near dynamic matter-dynamic dark energy equality. These considerations yield
\begin{eqnarray}\label{def3}
\dot{H}&=&-\frac{1}{2}\rho_{M}[(1+w_{M})+(1+w_{DDE})\frac{\rho_{DDE}}{\rho_{M}})].
\end{eqnarray} 
As in the previous section, we define a new parameter
\begin{eqnarray}\label{sysDD}
X_{ME}&=&\frac{\rho_{DDE}}{\rho_{M}}
\end{eqnarray}
The two dominant energy densities in this epoch are $\rho_{DDE}$ and $\rho_{M}$ ( the combination of dark matter and matter). 
\subsection{Deceleration parameter}
In terms of the Hubble parameter, the deceleration parameter takes the form
\begin{eqnarray}
-(q+1)&=&\frac{\dot{H}}{H^2},
\end{eqnarray} with $q>0$ signifying a decelerated expansion and $q<0$ an accelerated expansion. Using equations (\ref{def3}) and (\ref{Frid1}) we find,
\begin{eqnarray}\label{dec1}
-(1+q)&=&-\frac{3}{2}\bigg(\frac{1}{1+\chi_{ME}}\bigg)\bigg[ 1+(1+\omega_{DDE})\chi_{ME}\bigg],
\end{eqnarray} 
where we have neglected the ratio of all other density ratios other than that of the dynamic dark energy to combined matter. This is motivated by our intentions to study dynamics in the transition epoch where the two competing densities drive the universe's expansion. It will be noted, from equation (\ref{dec1}) that 
\begin{eqnarray}\label{dec2}
q&=&\frac{3}{2}\bigg[\frac{1+(1+\omega_{DDE})\chi_{ME}}{(1+\chi_{ME})}\bigg]-1.
\end{eqnarray} 
We see from equations (\ref{dec2}) that at equality, $\chi_{ME}=1$, that $\omega_{DDE}<-2/3$ implies accelerated expansion. On the other hand, if the dynamic dark energy mimics the cosmological constant $\omega_{DDE}=-1$, $X=0.5$ is sufficient to induce accelerated expansion. Investigations of the Pade model  \cite{Bouali23} seem to favour a positive deceleration parameter, inconsistency with observation results elsewhere.

In the fiducial model ($\kappa=0, \Omega_{0}=1$), observation data seem to prefer a value consistent with $\omega_{DE} = -0.94\pm 0.1$ \cite{Frie08}. 
 Lastly, we can express the $\omega_{DDE}$ as a function of the density ratio $X$ and the deceleration parameter $q$. This might help in reconstructing the EoS of dynamic dark energy if $q$ can be obtained from observation at a given density contrast. In particular,
\begin{eqnarray}\label{EoS}
\omega_{DDE}&=&\frac{(2q-1)(1+\chi_{ME})}{3\chi_{ME}},
\end{eqnarray} 
which is a formulation of EoS in terms of the deceleration parameter. This adds to the assortment of EoS already in literature such as Chevallier-Polarski-Linder \cite{Chevallier2001,Linder2008}, Barboza-Alcaniz \cite{Barboza2008} case and Low Correlation \cite{Wang2008} and their comparisons to observation data\cite{Bouali23,Alam04,Sahni04,Band22,Kundu23,Pade,Wei14,Rez17}.

\subsection{Transition dynamics}
We let $X=\chi_{ME}$ and $ Z=1/\rho_{M}$. Differentiating equations (\ref{sysDD}) and using equations (\ref{rhodots3}-\ref{rhodots5}) yields the following closed system of equations.
\begin{eqnarray}\label{sys2}
\dot{X}&=&3H(w_{M}-w_{DDE} )X-Z Q_{ME},\\
\dot{Z}&=&3H(1+w_{DM})Z,\\
\dot{H}&=&-\frac{1}{2Z}\left[1+w_{M}+X(1+w_{DDE})\right],
\end{eqnarray}  where term $Q_{rm}$ terms have been dropped from the system for the reason given above. We can also employ the ansatz $Q_{ME}=\psi X$. Setting the $w_{DM}=0=w_{M}$\cite{Kopp} yields the reduced system
\begin{eqnarray}\label{sys2}
\dot{X}&=&-3H\omega_{DDE}X- \psi XZ\\
\dot{Z}&=&3HZ\\
\dot{H}&=&-\frac{1}{2Z}\left[1+X(1+\omega_{DDE})\right],
\end{eqnarray} with $\omega_{DDE}$ given by equation (\ref{EoS}). 
This is the main system of equations in this study. 

\subsection{Results and analysis}
\begin{figure}[h!]
\centering
{\includegraphics[scale=0.45]{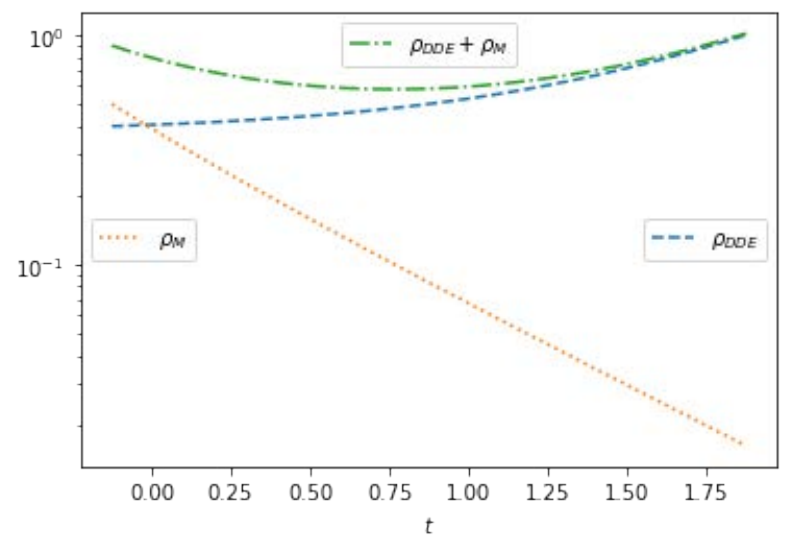} }
\caption{The dynamic energy density grows for $\omega_{DDE}=-0.7$. This has the potential to induce an accelerated expansion of the universe near $\chi_{ME}=1$ }.\label{fig6}
\end{figure}

\begin{figure}[h!]
\centering
{\includegraphics[scale=0.45]{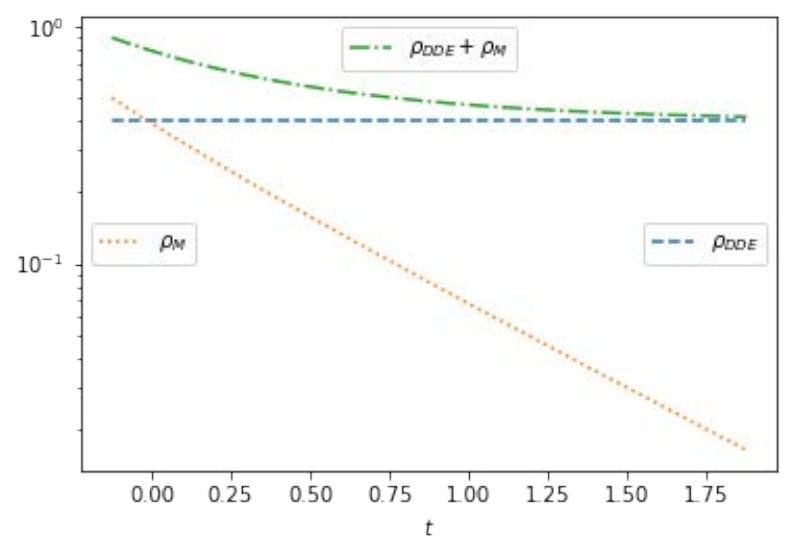}}
\caption{For $\omega_{DDE}=-1$, the dynamic dark energy is indistinguishable from $\Lambda$ and the resultant behaviour is what is expected if the cosmological constant drove the late time expansion.}\label{fig7}\end{figure}

\begin{figure}[h!]
\centering
{\includegraphics[scale=0.45]{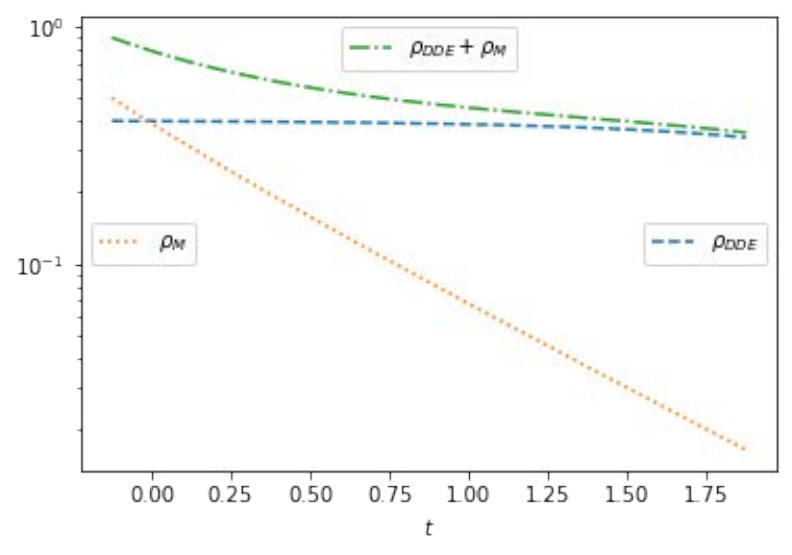}}
\caption{$w_{DM}=0$ and $w_{DDE}=0.5$. This delays the time to equality. }\label{fig8}\end{figure}

In this section, we consider different values of $q$ to understand the dynamics of the matter-to-dark energy transition. We have assumed that the acceleration is driven by dynamic dark energy $\rho_{DDE}$ rather than the cosmological constant (parametrised here by $\rho_{NDE}$). The dynamic dark energy considered is barotropic but with unknown EoS. We have reexpressed the equation of state in terms of the deceleration parameter and used this in finding the numerical system of equations (\ref{sys2}). 
The results are presented in two different formats. The first format looks at the behaviour of $\chi_{ME}, \rho_{M}$, and $\rho_{DDE}$ in the proximity of $\chi_{ME}=1$ for different values of $\omega_{DDE}$. The numerical solution results are given in Figures (\ref{fig6}, \ref{fig7} and \ref{fig8}).

\begin{figure}[h!]
\centering
{\includegraphics[scale=0.45]{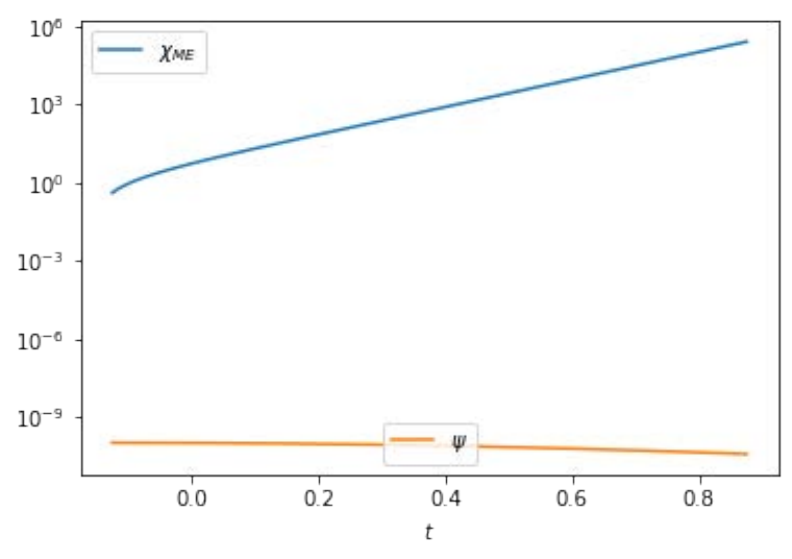}}
\caption{$\chi_{ME}$ grows with cosmic time. $\chi_{ME}=1$ is gven by $t_{eq}=0.$ The almost linear relation allows us to potentially fix $\chi_{ME}$ in equation (\ref{EoS}). } \label{fig9}\end{figure}

\begin{figure}[h!]
\centering
{\includegraphics[scale=0.45]{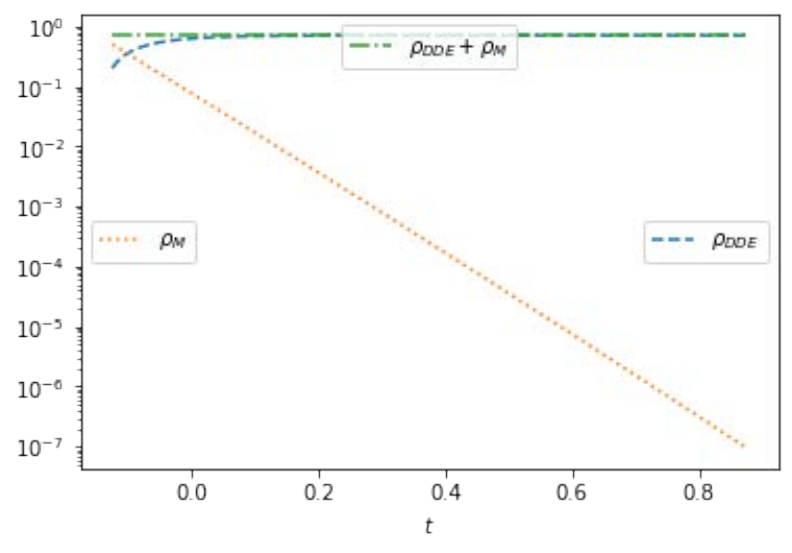}}
\caption{The accelerated expansion case for $q=-1$. One can reconstruct the $\omega_{DDE}$ using equation (\ref{EoS}) leading to EoS of $\omega_{DDE}=-2$ that lies in the phantom \cite{LiZha11} regime at $\chi_{ME}=1$.}\label{fig10} \end{figure}

\begin{figure}[h!]
\centering
{\includegraphics[scale=0.45]{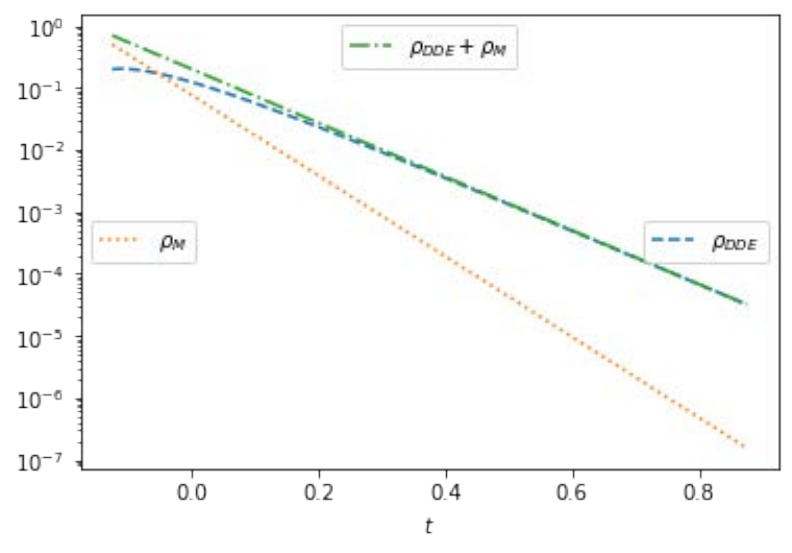}}
\caption{ This is the case of vanishing deceleration, $q=0$. The dynamic dark energy decays with expansion to mimic baryonic matter. It is equivalent to $\omega_{DDE}=-2/3.$}\label{fig11}\end{figure}

\begin{figure}[h!]
\centering
{\includegraphics[scale=0.45]{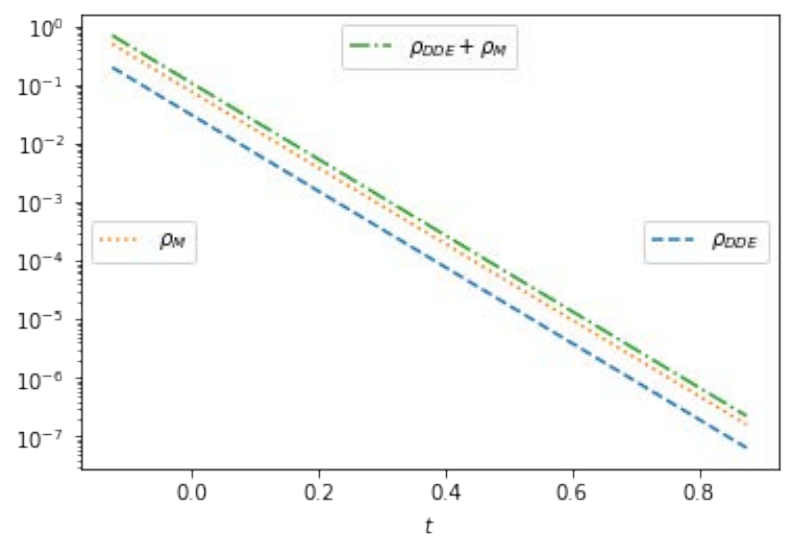}}
\caption{$q=0.5$ is the sweet spot for no cross-over between matter and dynamic dark energy i.e. the dynamic energy remains subdominant.}\label{fig12}\end{figure}

The second format looks at the same parameter and properties but for different values of the deceleration parameter. The results are given in Figures (\ref{fig9},\ref{fig10},\ref{fig11} and \ref{fig12}).
\begin{figure}[h!]
\centering
{\includegraphics[scale=0.45]{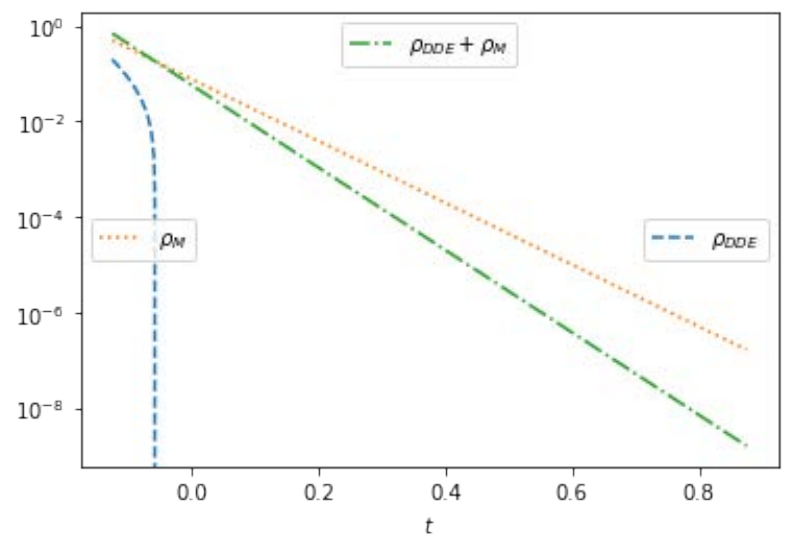}}
\caption{The case of $q=1$, see the dynamic dark energy decays exponentially fast without a cross-over. }\label{fig13}\end{figure}

\section{\label{DC}Discussion and Conclusion}
One of the main results of this paper is a new equation of state for dynamic dark energy given in terms of density contrast between two succeeding dominant energy densities. I.e.
\begin{eqnarray}
\omega_{DDE}&=&\frac{(2q-1)(1+\chi_{ME})}{3\chi_{ME}},
    \end{eqnarray} where $q$ is the deceleration parameter and $\chi_{ME}$ is the ratio of dynamic dark matter energy density to that of matter. The uncertainty in measure of the Hubble parameter implies uncertainty in obtaining the exact value of $q$ and consequently the $\omega_{DDE}$. Nevertheless, the numerical solution of the coupled system of equations involving interacting energy densities indicates a preference for $\omega_{DDE}<-2/3$ for the induction of an accelerated expansion at cross-over; $\chi_{ME}=1.$  The $-1<\omega_{DDE}<-2/3$ corresponds to EoS in the quintessence regime\cite{Caldwell}.

A recent analysis of various dynamic dark energy EoS against a collection of data sets \cite {Rahman20}, found that flat $\Lambda$-CDM and WCDM models using the BIC criteria provide a greater agreement with the data sets used. This does rule out dynamic dark energy as a possible explanation for cosmic acceleration. It, nevertheless, places stringent bounds on the EoS for suitable dynamic dark energy models. The current debate on Hubble tension motivates further investigations into EoS dynamic dark energy.

Acknowledgement:
The author thanks the University of Cape Town's NGP for financial support. The original version of this article appears online in arXiv:2002.08875.

Appendix

\end{document}